\documentclass[journal]{IEEEtran}
\usepackage{graphicx}
\usepackage{subfigure}
\usepackage{multirow}
\usepackage{amsmath}%
\usepackage{MnSymbol}%
\usepackage{wasysym}%
\usepackage{gensymb}

\usepackage{atbegshi}
\AtBeginDocument{\AtBeginShipoutNext{\AtBeginShipoutDiscard}}

\begin{document}
\title{Testing sTGC with small angle wire edges for the ATLAS New Small Wheel Muon Detector Upgrade}
%
%

\author{Itamar Roth,
        Amit Klier
        and~Ehud Duchovni}
\thanks{Itamar Roth, Amit Klier and Ehud Duchovni are with the Weizmann Institute of Science, 234 Herzl St., Rehovot 7610001, Israel (e-mail: itamar.roth@weizmann.ac.il).}%
\maketitle
\pagestyle{empty}
\thispagestyle{empty}

\begin{abstract}
The LHC upgrade scheduled for 2018 is expected to significantly increase the accelerator's luminosity, and as a result the radiation background rates in the ATLAS Muon Spectrometer will increase too.
Some of its components will have to be replaced in order to cope with these high rates.
Newly designed small-strip Thin Gap chambers (sTGC) will replace them at the small wheel region.
One of the differences between the sTGC and the currently used TGC is the alignment of the wires along the azimuthal direction.
As a result, the outermost wires approach the detector's edge with a small angle.
Such a configuration may be a cause for various problems.
Two small dedicated chambers were built and tested in order to study possible edge effects that may arise from the new configuration. The sTGC appears to be stable and no spark have been observed, yet some differences in the detector response near the edge is seen and further studies should be carried out.
\end{abstract}


\section{Introduction}
\begin{figure}[!t]
\centering
\includegraphics[width=3.5in]{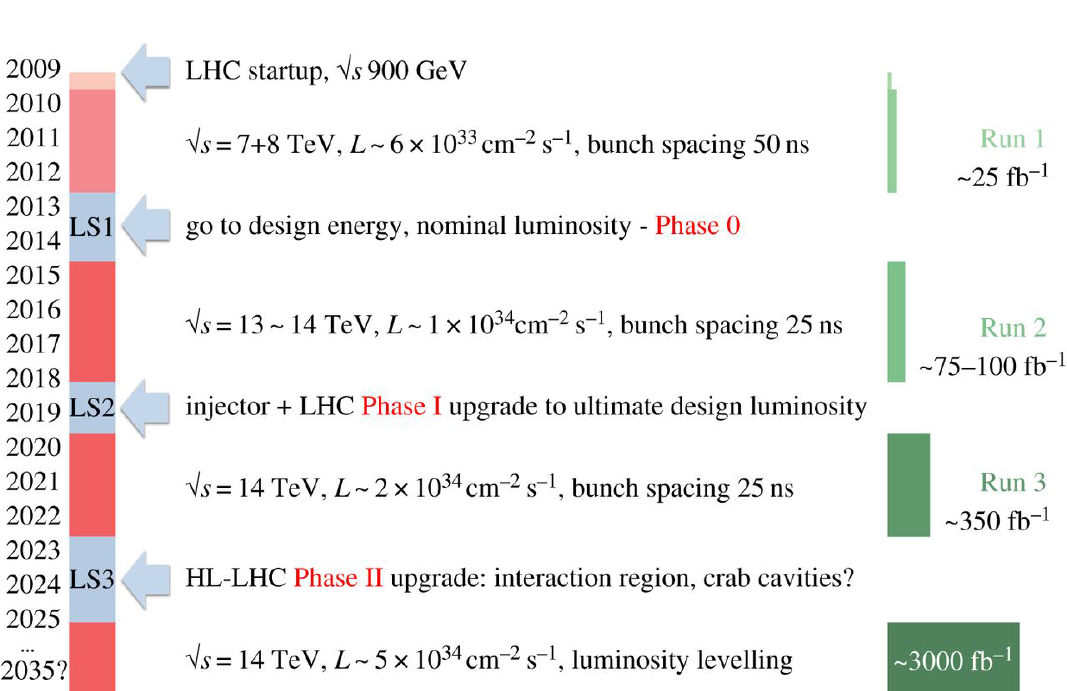}
\caption{The LHC timeline.}
\label{fig:lhc_timeline}
\end{figure}

\IEEEPARstart{T}{he} LHC is expected to be upgraded to the HL-LHC (High Luminosity LHC)~\cite{Tricomi:2008zz} in several phases with the goal of obtaining an instantaneous luminosity of 5 $\times~10^{34}$ cm$^{-2}$s$^{-1}$ at a center of mass energy of 14 TeV, as described in Figure~\ref{fig:lhc_timeline}.
After the long shutdown scheduled for 2018 (LS2), the LHC luminosity will be increased by a factor of two compared to the upcoming 2015 run. The collider is then expected to reach a luminosity of 2 $\times~10^{34}$ cm$^{-2}$s$^{-1}$.
The ATLAS detector~\cite{Aad:2008zzm} will also have to be upgraded in stages in order to cope with the higher collision rates and the elevated radiation background rates.

\subsection{The New Small Wheel}\label{sec:TGC_intro}

The SW (Small Wheel) is the innermost part of the ATLAS Muon Spectrometer~\cite{ATLAS:1997ad}. It covers the pseudo-rapidity range of $1.3 < |\eta| < 2.7$.
The SW consists of MDT (Monitored Drift Tube) and CSC (Cathode Strip Chamber) systems that are used for muon tracking. It also contains a TGC (Thin Gap Chamber) system for triggering purposes.
Both MDT and CSC will be unable to cope with the elevated radiation levels expected at the HL-LHC.
Hence, the ATLAS collaboration decided to replace the present SW with a nSW (new Small Wheel)~\cite{Kawamoto:1552862}.
The nSW will employ a new type of small-strip, Thin Gap Chambers (sTGC)~\cite{GERBAUDO:2014wqa} and Micromegas detectors~\cite{Zibell:2014fea} for both triggering and precision tracking.

\subsection{New Thin Gap Chambers Design}\label{sec:TGC_design}

The TGC is a multi-wire chamber with 50~$\mu$m diameter gold-plated tungsten wires, forming the anode plane. Resistive carbon, coating the FR4 walls, serve as a cathode.
The operational gas is a mixture of CO$_2$ and n-pentane (C$_5$H$_{12}$) with a ratio of 55:45 at atmospheric pressure.
The anode to cathode spacing is 1.4~mm and the wire to wire spacing is 1.8~mm. The operating voltage of this device is about 3~kV.
A total of $\sim$ 3600 TGCs were installed in ATLAS.
They are arranged in six big wheels and two small ones, divided into wedges, and therefore have a trapezoid shape.

\begin{figure}[!t]
\centering
\includegraphics[width=3.5in]{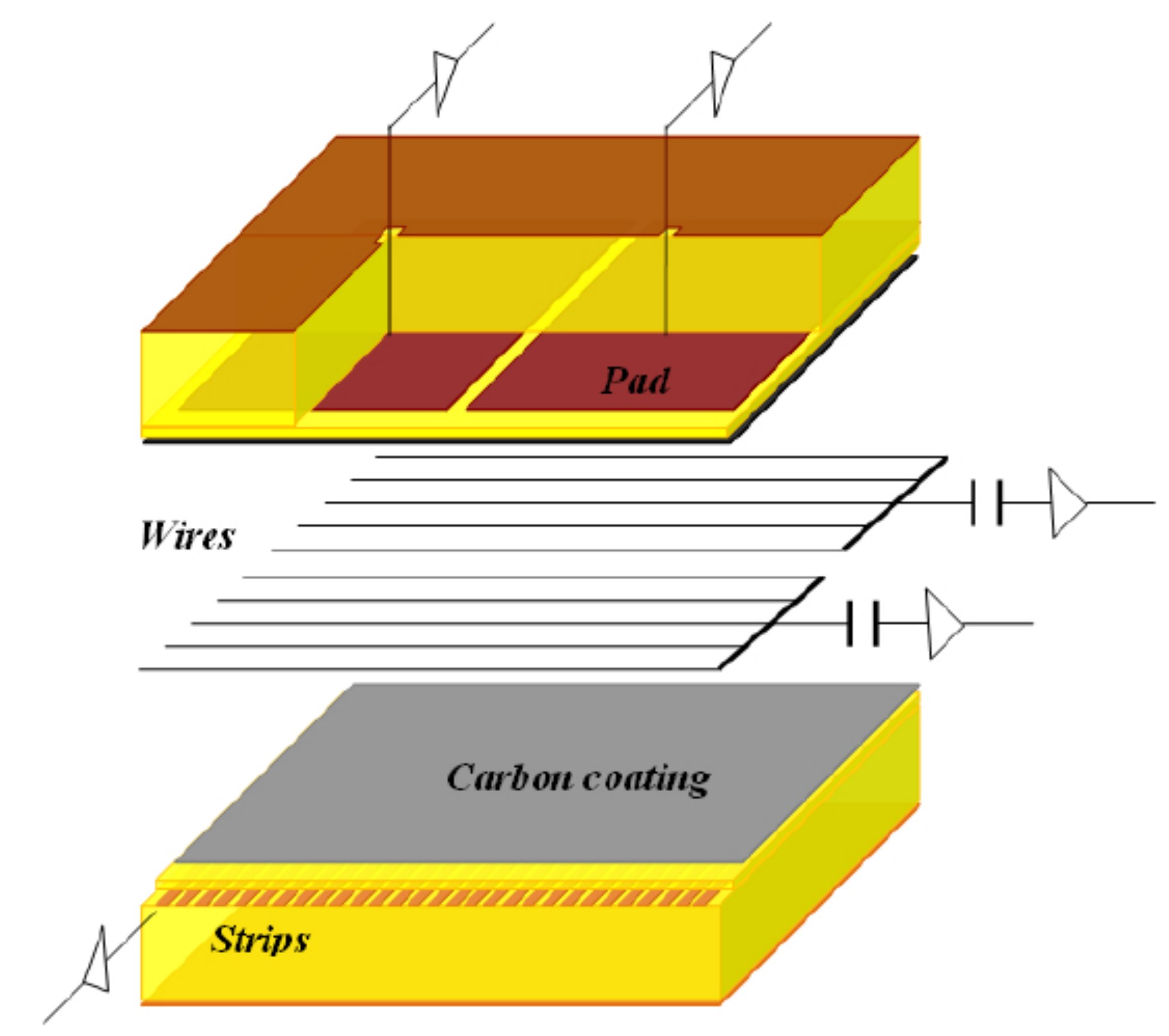}
\caption{A schematic cross-section of a sTGC detector.}
\label{fig:TGC_scheme}
\end{figure}

The sTGC chambers are a variant of the ATLAS TGC ones.
Each chamber is equipped with a series of pad readouts for the first level trigger, strip readout for high precision tracking, and wire readout for the determination of the second coordinate.
To decrease charge accumulation on the cathode when the chamber operates at high rate, the cathode surface resistivity is reduced from $\sim \rm{M\Omega/\Box}$, currently used at the ATLAS TGC, to $\sim 100~\rm{k\Omega/\Box}$. A schematic view of the sTGC is shown in Figure~\ref{fig:TGC_scheme}.
The main sTGC parameters are listed in Table~\ref{table:tgc_parameters}.

\begin{table}[!tbh]
\begin{center}
\begin{tabular}{c|c}
\hline\hline
\rule{0pt}{2ex} sTGC geometry & Value \\
\hline
\rule{0pt}{2ex} Wire-carbon gap & 1.4~mm \\[1ex]
\rule{0pt}{2ex} Wire-wire space & 1.8~mm \\ [1ex]
\rule{0pt}{2ex} Strip-carbon gap & 0.1~mm \\[1ex]
\rule{0pt}{2ex} Strip pitch & 3.2~mm \\[1ex]
\rule{0pt}{2ex} Inter-strip gap & 0.5~mm \\[1ex]
\rule{0pt}{2ex} Cathode plate resistivity & $100~\rm{k\Omega/\Box}$\\
\hline\hline
\end{tabular}
\caption{sTGC parameters.}
\label{table:tgc_parameters}
\end{center}
\end{table}

\subsection{Motivation}\label{sec:TGC_motivation}

\begin{figure}[!t]
\centering
\includegraphics[width=3.5in]{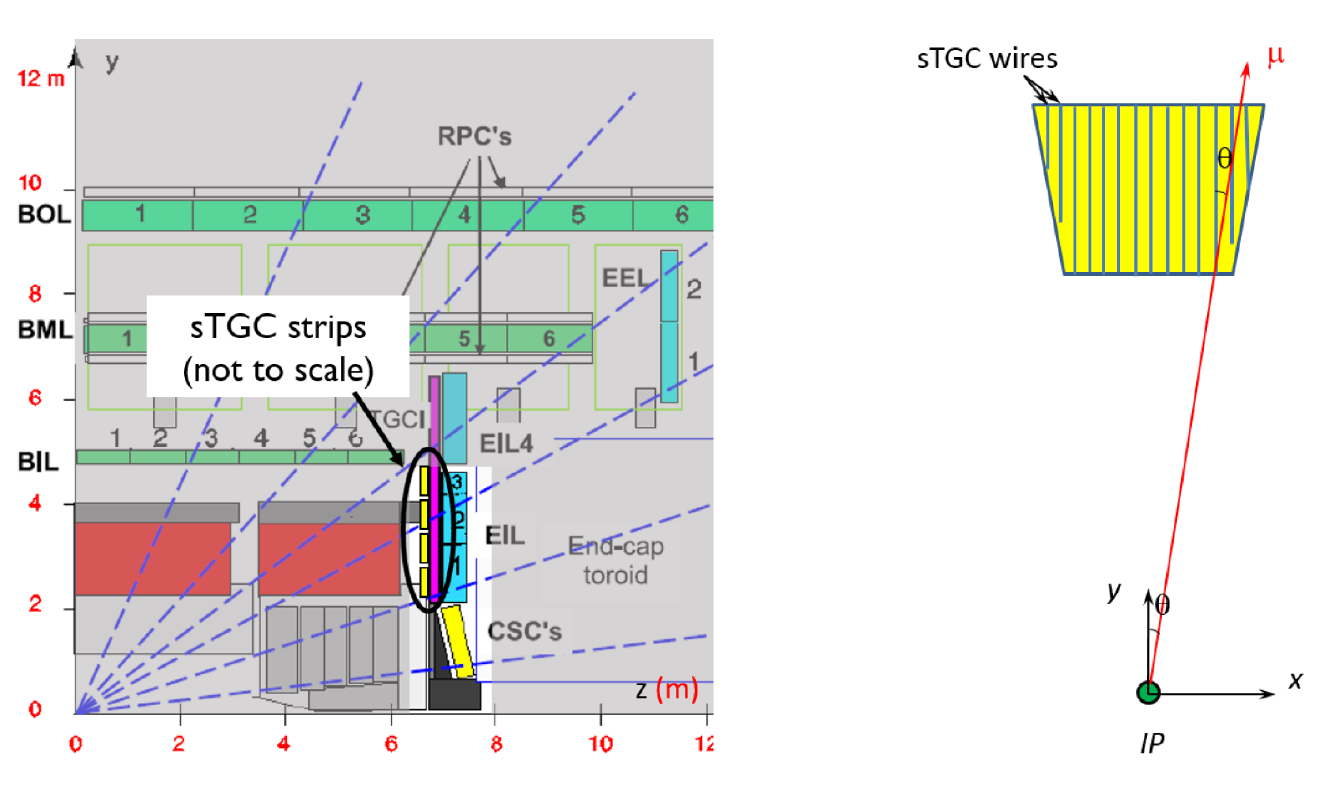}
\caption{The orientation of sTGC strips in nSW as viewed in the $yz$ direction (left) and the wires direction viewed towards the $z$ direction (right).}
\label{fig:TGC_orientation}
\end{figure}

While the present ATLAS TGC is primarily used for triggering, the sTGC is also expected to provide precision tracking in the radial direction, namely, in the direction that determines the momentum of the track.
For triggering, the sTGC detectors are required to identify each muon's bunch crossing and to measure its trajectory with an angular resolution of less than 1~mrad for $1.3 < |\eta| < 2.7$.
For tracking, the chambers are required to have a position resolution better than 100~$\mu$m at impact angles up to $30\degree$. 
To achieve this goal, the sTGC detectors are equipped with fine-pitch strips in $\phi$. A center of gravity (COG) algorithm determine the hit location along the wires which are perpendicular to the strip direction and, therefore, are approximately aligned along the azimuthal direction (perpendicular to the trapezoid bases) as illustrated in Figure~\ref{fig:TGC_orientation}.
This differs from the present TGC detectors, in which the wires are strung along the $\phi$ direction (parallel to the bases).

The new structure results in a small angle between the wires and the detector edge at the trapezoid legs. This proximity may give rise to unwanted edge effects, such as sparks or a disturbance of the electric field, which may affect the detector's efficiency and spatial resolution.
The goal of this paper is to study the severity of such effects.

\section{Methods}\label{sec:TGC_methods}

\subsection{Experimental Setup}\label{sec:TGC_setup}

\begin{figure}[!t]
\centering
\includegraphics[width=3.5in]{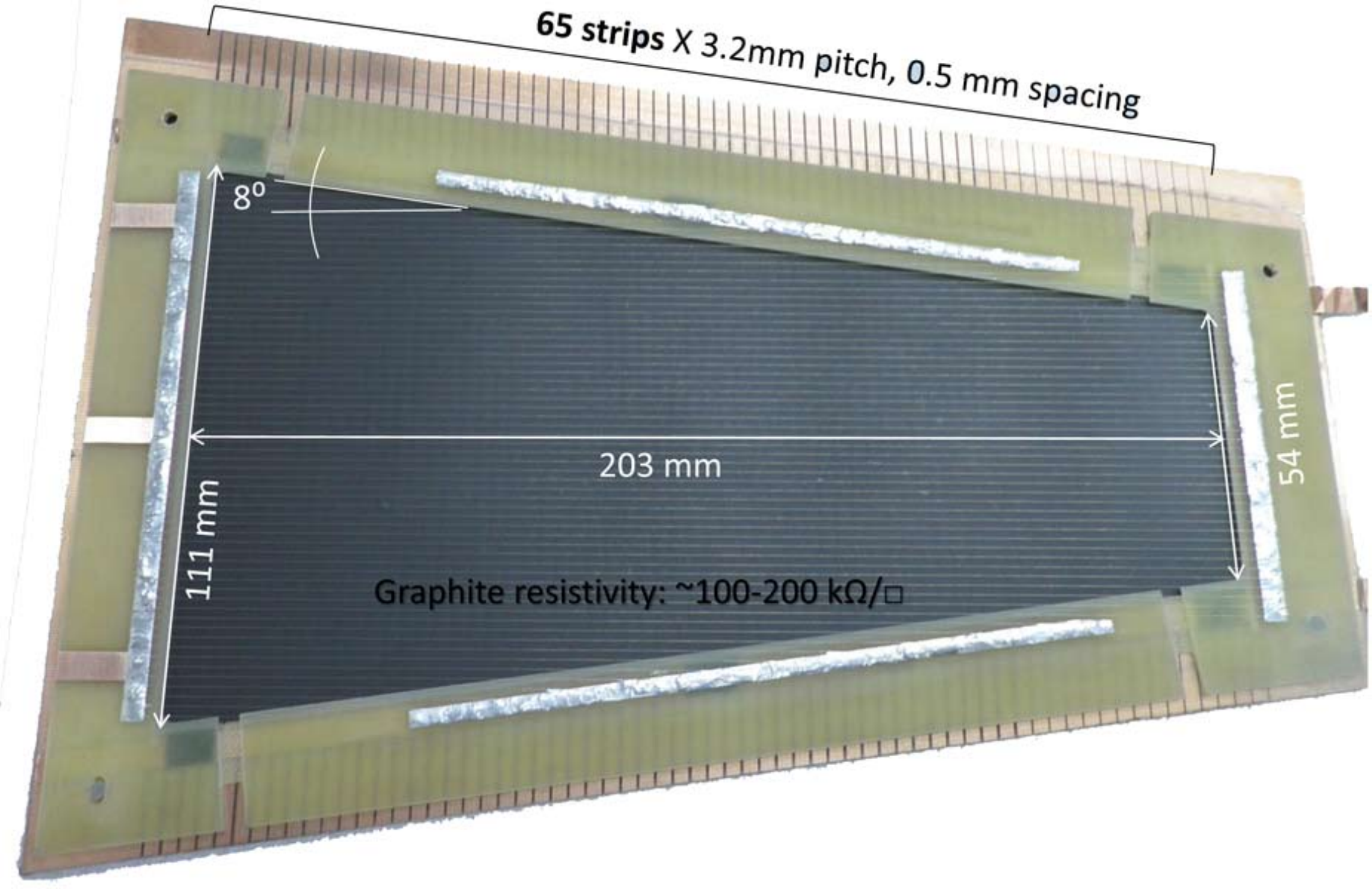}
\caption{An open sTGC prototype and its main features.}
\label{fig:TGC_pro}
\end{figure}

\begin{figure}[!t]
\centering
\includegraphics[width=3.5in]{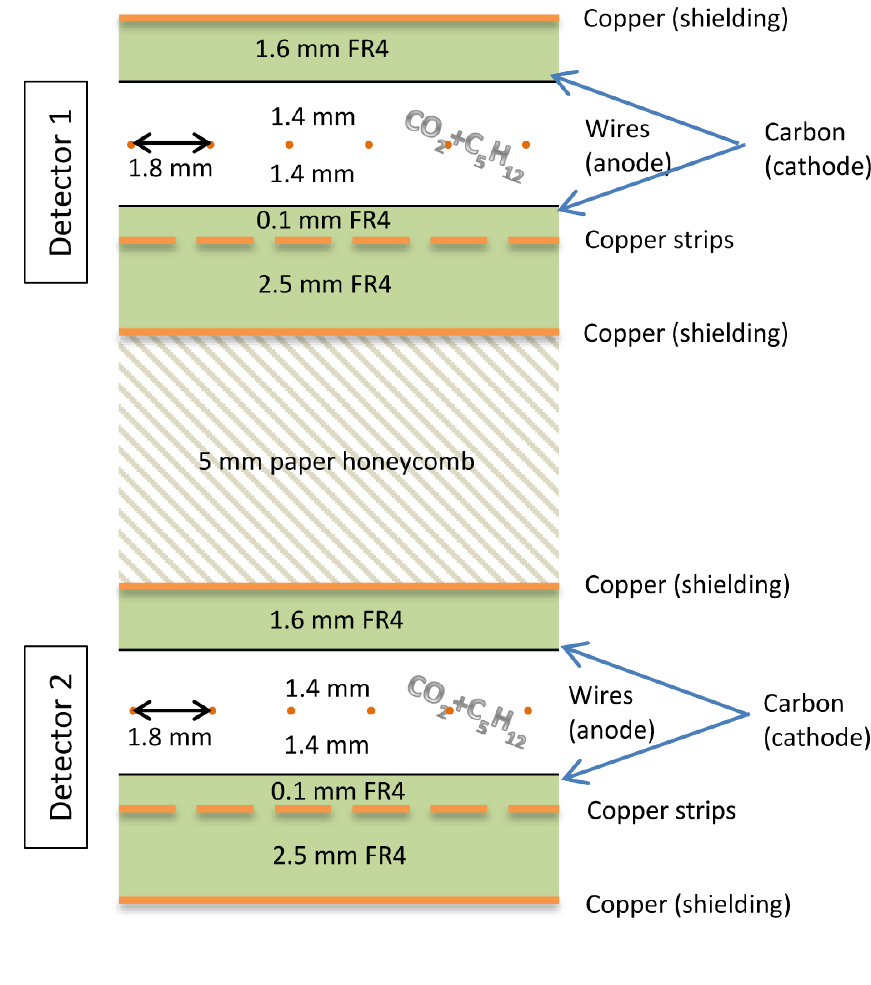}
\caption{Cross-section of the sTGC prototype doublet (not to scale).}
\label{fig:TGC_cross_section}
\end{figure}

A small prototype sTGC doublet of approximately 0.1$~\times$~0.2~$\rm{m}^2$ in area consisting of two parallel-placed chambers was constructed for the purpose of the present tests.
The chambers are equipped with strips on one side, and have no pads on the other.
The main features of the prototype are shown in Figure~\ref{fig:TGC_pro}.
The angle between the  wires and the detector edge at the trapezoid legs is  $8\degree$.
The distance between the two wire planes is 12~mm, including a 5~mm paper honeycomb layer separating the chambers.
An illustrated cross-section of the doublet is shown in Figure~\ref{fig:TGC_cross_section}. Each chamber is equipped with 65 strips (3.2~mm pitch: 2.7~mm strip width and 0.5~mm gap), out of which 32 are connected to readout electronics. 
The actual sTGC chamber is expected to be larger ( $\sim 1~\rm{m}^2$ on average) and the trapezoid angles are expected to be either $6\degree$ or $14\degree$.

\begin{figure}[!t]
\centering
\includegraphics[width=3.5in]{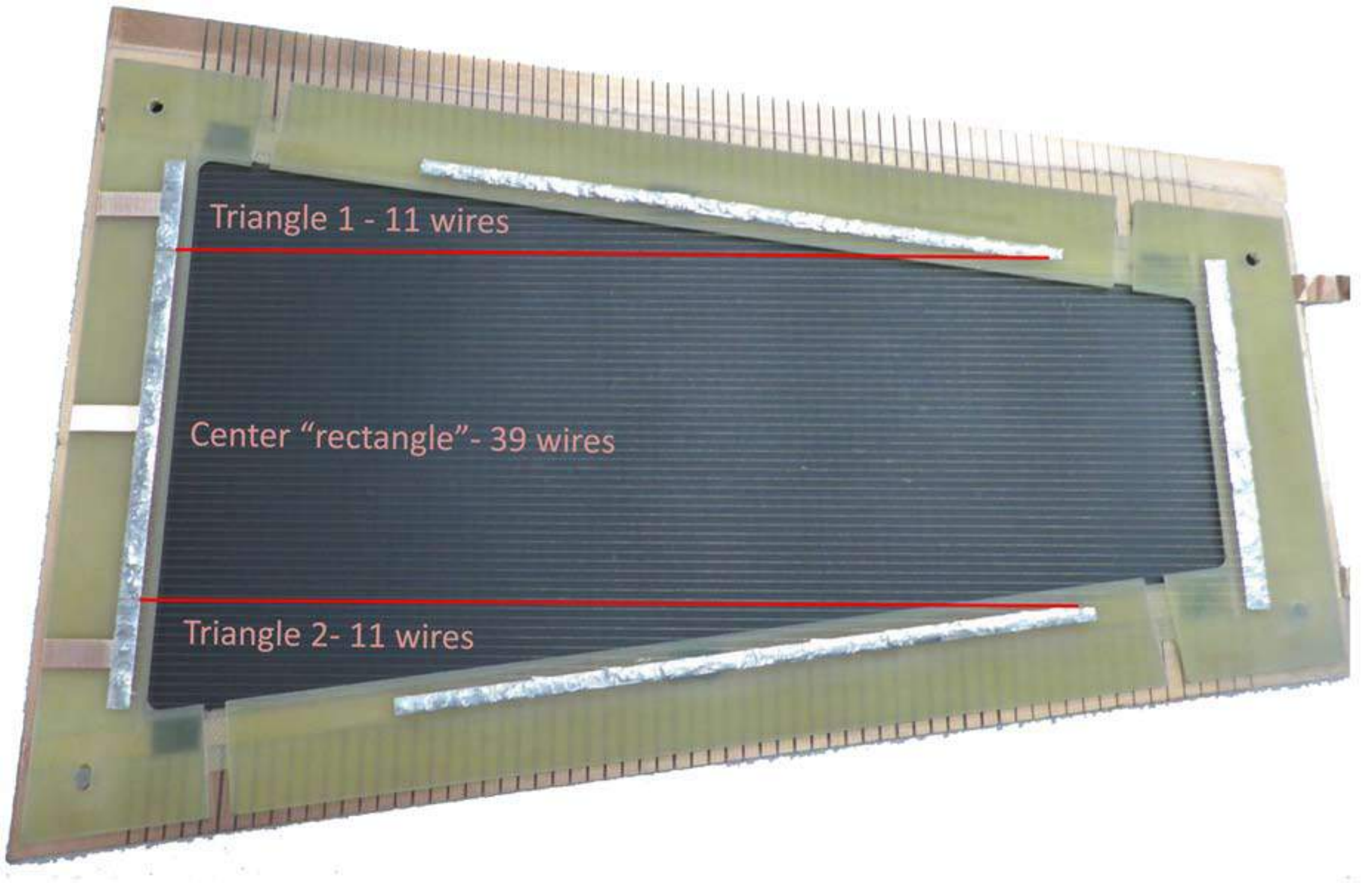}
\caption{The sTGC prototype and its division into three geometrical wire groups.}
\label{fig:TGC_Geometry}
\end{figure}

In order to study the edge effects, the edge wires and the central wires were grouped independently. Figure~\ref{fig:TGC_Geometry} shows this division into three wire groups: a central rectangle and two side triangles. 3 out of the 32 readout channels in each chamber were connected to these wire groups to provide the "geometrical coordinate" for the following analysis.

\subsection{Initial Tests}\label{sec:TGC_initial}

Beta electrons from a radioactive $^{90}$Sr source were used in order to study the turn-on curve and to set the optimal working voltage for the tested chambers.
The efficiency of each chamber was checked separately, and found to be higher than 99$\%$ for operating voltage of 2.93~kV for one chamber and 2.9~kV for the other. The actual sTGC system is expected to operate at about this voltage. 
No sparks near the trapezoid legs were observed even at 20$\%$ higher operating voltage, namely at 3.5~kV.

\subsection{Position Measurement and Event Selection}\label{sec:evnt_selection}

The main test was done using cosmic particles. The readout system was triggered when wire signals from both chambers coincided within a 30~ns time-window. The wire signals were digitized by the ASD (Amplifier-Shaper-Discriminator) of the same type as those currently used by the ATLAS TGC, while the strip signals were amplified and shaped using the analog part of the ASD electronics.
The charge collected within a 100~ns time-window in each strip was digitized and recorded using VME CAEN V792 modules. 

\begin{figure}[!t]
\centering
\includegraphics[width=3.5in]{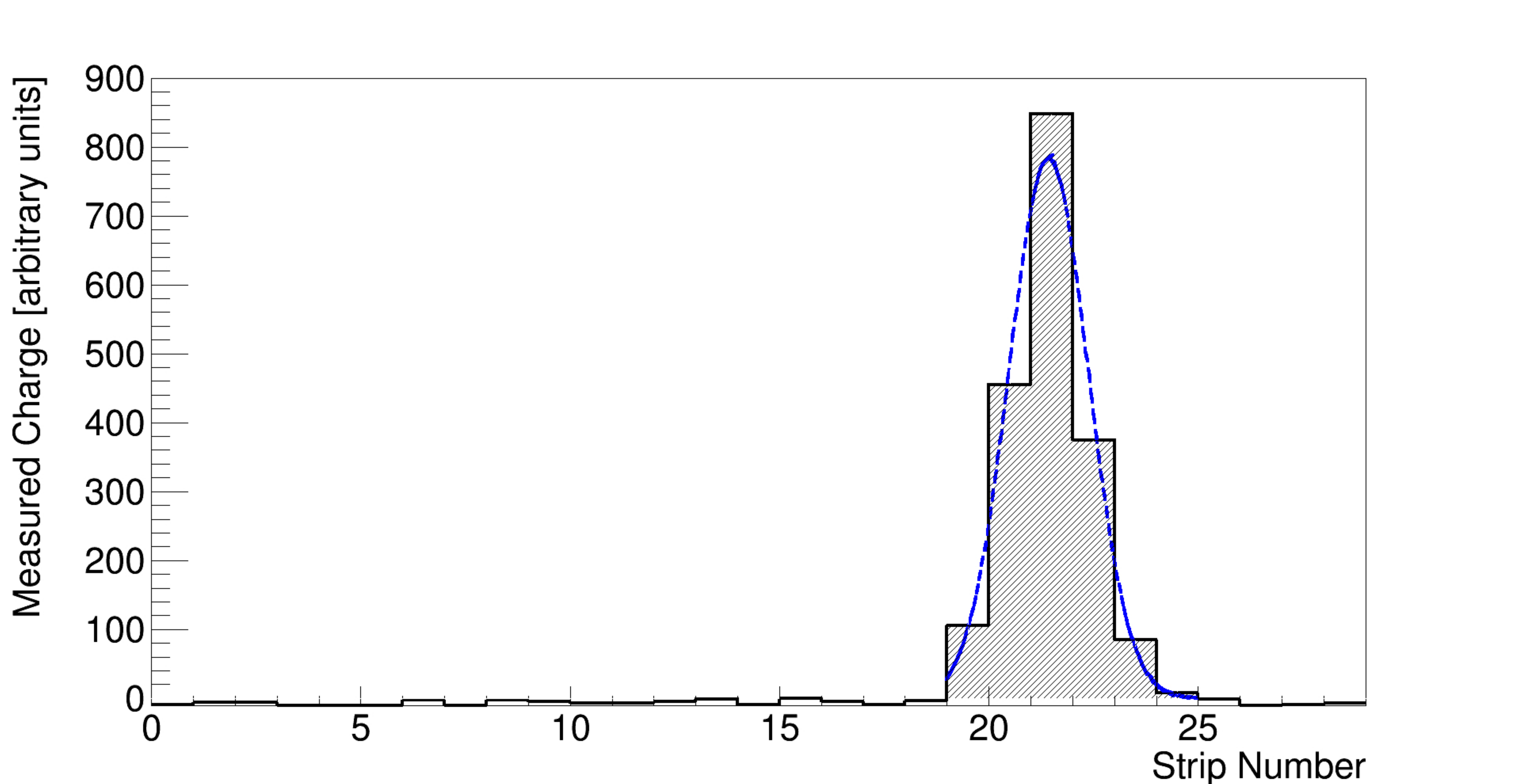}
\caption{A typical event profile, including  a Gaussian fit (blue dashed line). The filled area is the "raw area" of the signal, which corresponds to the integral of the measured charge.}
\label{fig:signal_example}
\end{figure}

The electron avalanche that results from the gas ionization due to a particle that hits the detector, is collected on the wire and induced on the strips.
The induced charge is distributed on several strips around the hit position, with additional smearing due to the carbon layer. 
The hit position is measured by fitting a Gaussian to the strip's signal strengths.
This method was used at previous tests~\cite{Amram:2010ax, Benhammou:2011kg} and showed that a single layer sTGC can provide a spatial resolution of the order of 50~$\mu$m.
A typical event profile recorded by the sTGC prototype is shown in Figure~\ref{fig:signal_example}.
Low position values correspond to the long base of the trapezoid and high values to the short base.

First, several quality cuts are applied:

\begin{itemize}
\item[1]	The event amplitude (maximal charge collected on a strip) has to be above a certain threshold (3 times the standard deviation of the pedestal level) -- to suppress noise;
\item[2]	The strip where the collected charge is maximal should not be either first or the last strip (namely, those who has read-out strips only on one side) -- to select well-contained signals;
\item[3]	Events with a strip in charge overflow are rejected;
\item[4]	The Gaussian fit has to converge with a small $\chi^{2}$ / n.d.o.f. -- to assure good reconstruction;
\item[5]	The Gaussian fit standard deviation ($\sigma$) is less than two strips wide -- to remove delta electrons and other non-minimum ionizing particles.
\end{itemize}

About 50\% of the recorded events survived these quality cuts.

\section{Results}\label{sec:TGC_results}

\begin{figure}[!t]
\centering
\subfigure[]{
  \includegraphics[width=3.5in]{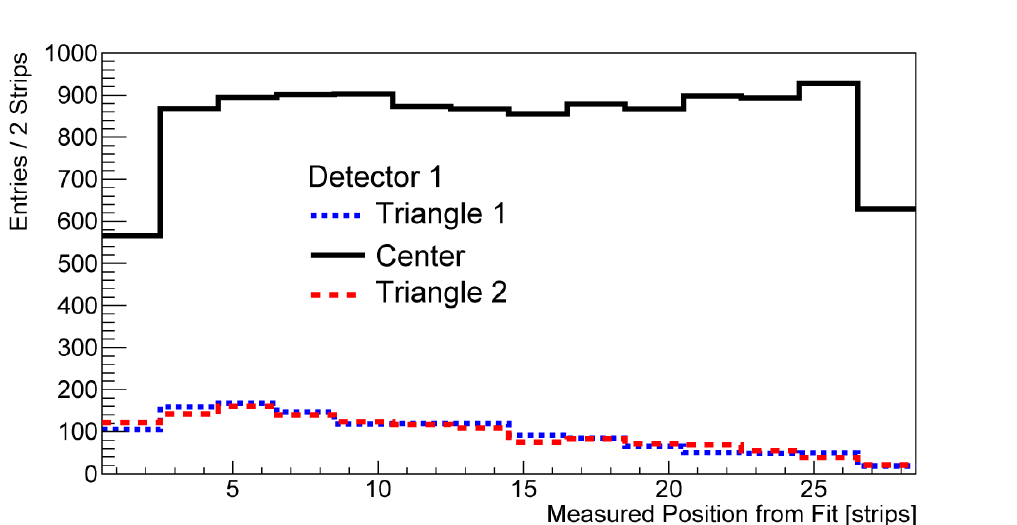}
}
\subfigure[]{
  \includegraphics[width=3.5in]{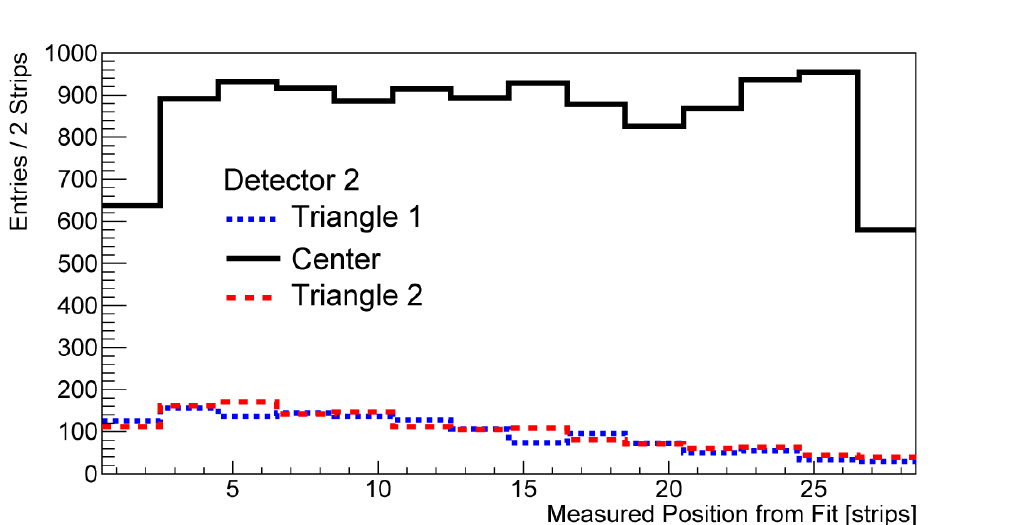}
}
\caption{The measured hit position in the three chamber regions. The distributions reflect the geometrical shape of each region in detector 1 (a) and detector 2 (b).}
\label{fig:TGC_Geometry2}
\end{figure}

Figure~\ref{fig:TGC_Geometry2} shows the position distribution of signal events.
It is shown separately for the three wire groups defined in Section~\ref{sec:TGC_setup}. 
the trapezoidal shape of the detector is well reflected in the measured density distribution, as there are more hits at long wires than at short ones. The relatively low number of events in the side triangles, compared to the central rectangle, corresponds to their relative area.
The small number of entries in the two extreme bins is due to the second quality cut presented in Section~\ref{sec:evnt_selection}.

\begin{figure}[!t]
\centering
\subfigure[]{
  \includegraphics[width=3.5in]{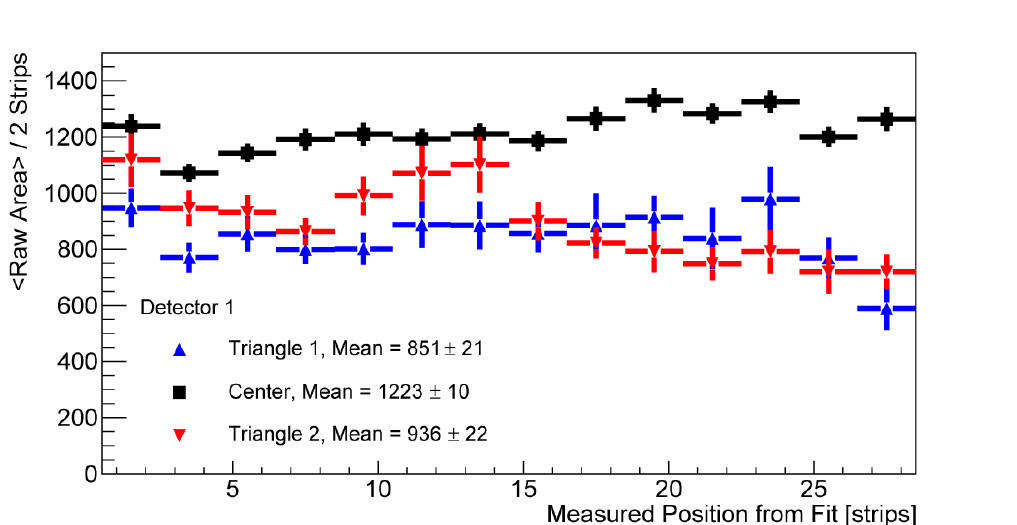}
}
\subfigure[]{
  \includegraphics[width=3.5in]{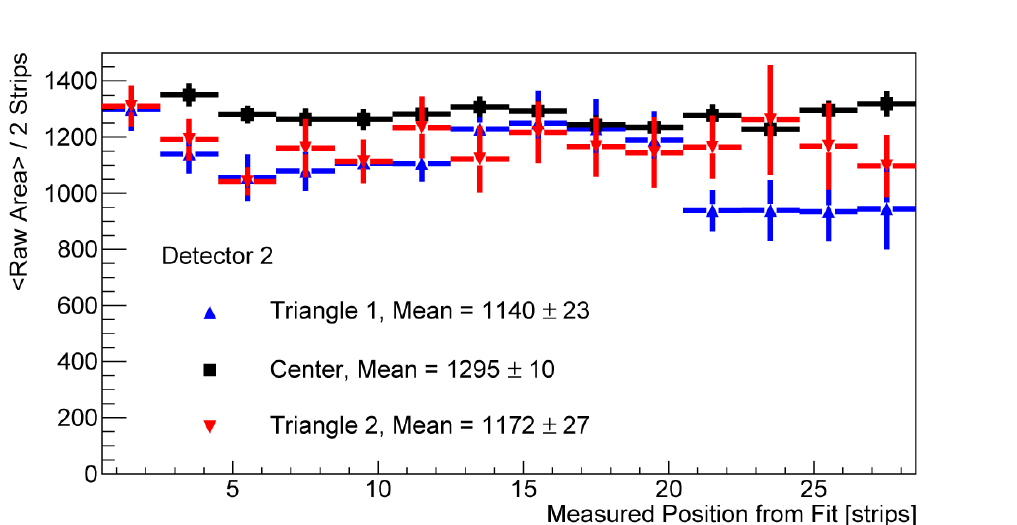}
}
\caption{The raw area (proportional to the measured charge) as a function of the measured hit position in the three wire groups in detector 1 (a) and detector 2 (b).}
\label{fig:avg_area_vs_pos}
\end{figure}

\begin{figure}[!t]
\centering
\subfigure[]{
  \includegraphics[width=3.5in]{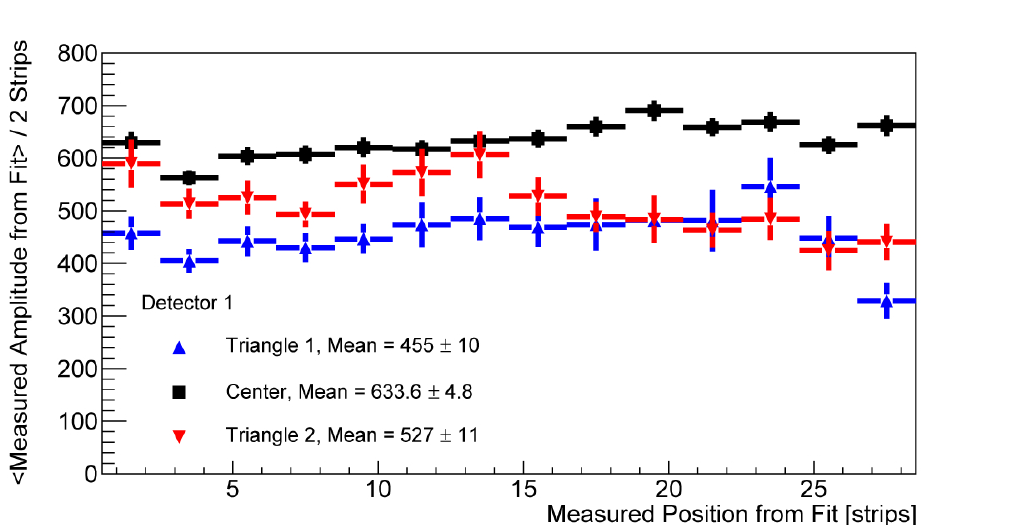}
}
\subfigure[]{
  \includegraphics[width=3.5in]{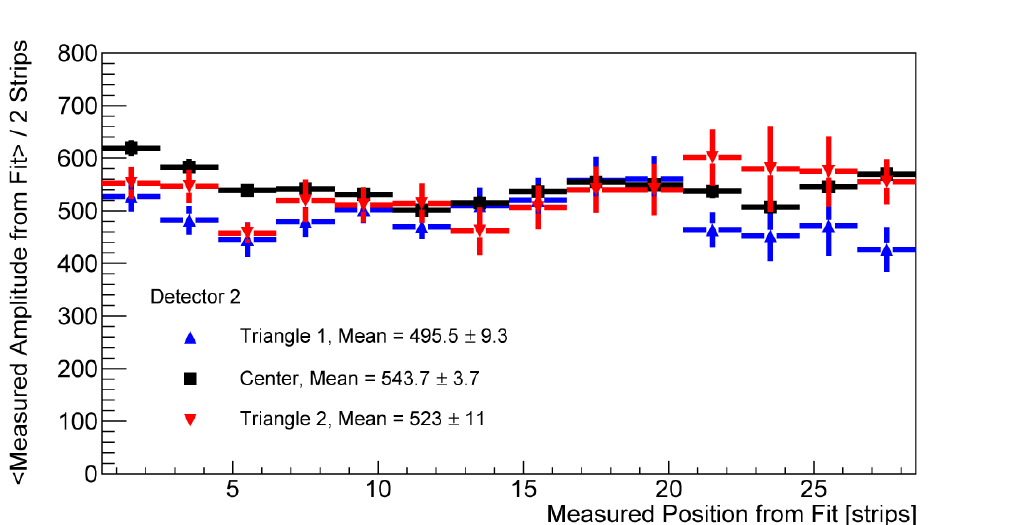}
}
\caption{The amplitude as deduced from the Gaussian fit  as a function of the measured hit position in the three wire groups in detector 1 (a) and detector 2 (b).}
\label{fig:avg_amp_vs_pos}
\end{figure}

Figure~\ref{fig:avg_area_vs_pos} shows the average "raw area" of the signal as a function of the measured hit position along the wire.
The raw area of the signal corresponds to the integral of the measured charge, as shown in Figure~\ref{fig:signal_example}.
Figure~\ref{fig:avg_amp_vs_pos} shows the average amplitude of the Gaussian fit as a function of the measured hit position.
Table~\ref{tab1} summarizes the differences between the side triangles and the central rectangle for the measured parameters.
The signals in the side triangles are smaller (both in area and in amplitude) than the ones in the central rectangle.
The difference between the triangles of the two chambers may be explained by differences in the construction of the two detectors, specifically, non-uniform spacing between the detector walls and the wire plane, even within the same chamber.

\begin{table}[!t]
\begin{center}
\begin{tabular}{c|ccc}\hline\hline
\rule{0pt}{2ex}		         & Ratio                                            & Raw area                & Fit amplitude\\\hline
\multirow{2}{*}{Detector 1} & \rule{0pt}{3ex} Triangle 1 / Center & $69.5\%\pm1.8\%$ &  $71.7\%\pm1.7\%$\\[1ex]
					  & Triangle 2 / Center                        & $76.5\%\pm1.9\%$ &  $83.1\%\pm1.8\%$\\[1ex]\hline
\multirow{2}{*}{Detector 2} & \rule{0pt}{3ex} Triangle 1 / Center & $88.0\%\pm1.9\%$ &  $91.1\%\pm1.8\%$\\[1ex]
					 & Triangle 2 / Center                         & $90.5\%\pm2.2\%$ &  $96.2\%\pm2.1$\%\\[1ex]
\hline\hline
\end{tabular}
\end{center}
\caption{The mean ratio between the side triangles and the central rectangle for the different measured parameters.}  
\label{tab1}
\end{table}

\begin{figure}[!t]
\centering
\subfigure[]{
  \includegraphics[width=3.5in]{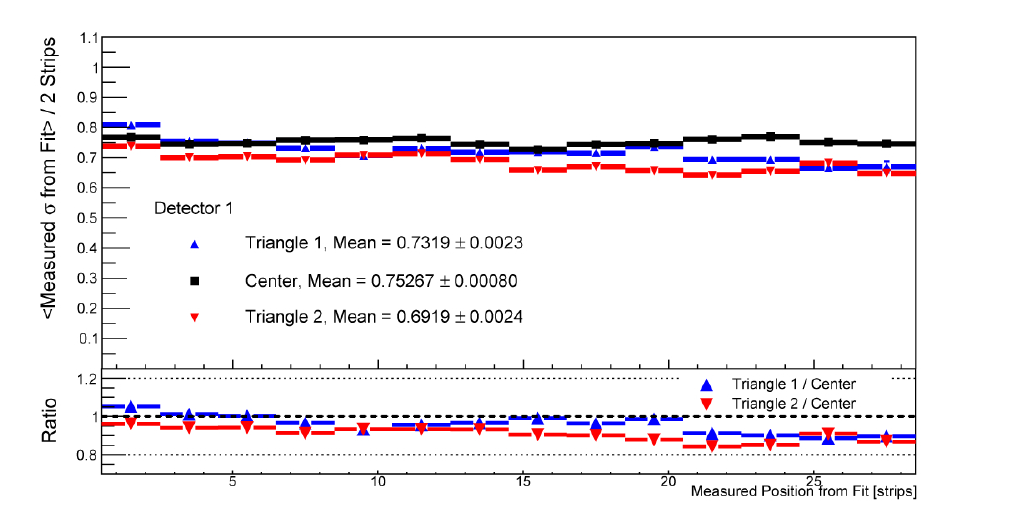}
}
\subfigure[]{
  \includegraphics[width=3.5in]{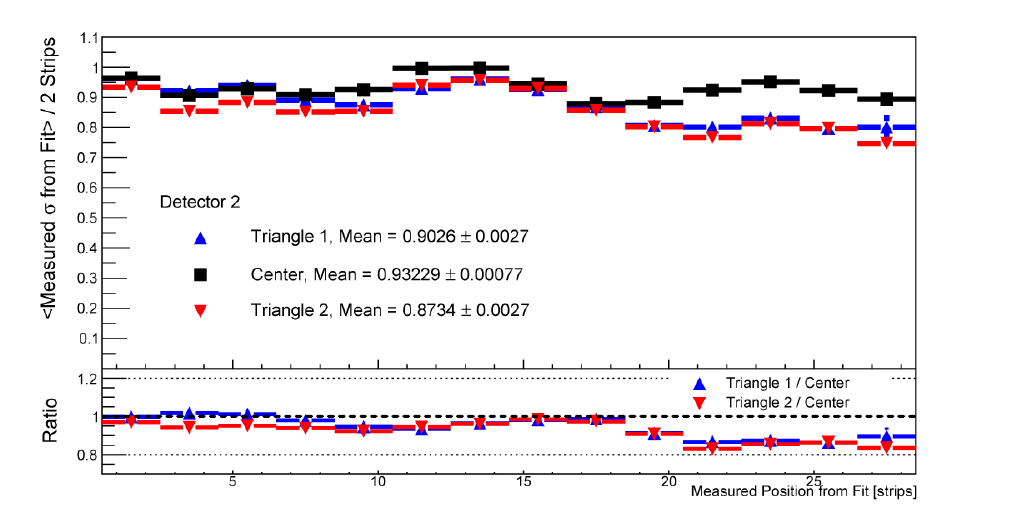}
}
\caption{The width of the Gaussian fit  as a function of the measured hit position in the three wire groups in detector 1 (a) and detector 2 (b).}
\label{fig:avg_sig_vs_pos}
\end{figure}

Figure~\ref{fig:avg_sig_vs_pos} shows the width ($\sigma$) of the Gaussian fit as a function of the measured hit position.
As can be seen in the bottom panels, $\sigma$ decreases by $\sim ~15\%$ as the triangles narrow.
This tendency is found in all triangles in both chambers.
It is possible that the structure featured in the middle of Figure~\ref{fig:avg_sig_vs_pos}(b) is due to the non-uniformities of the graphite in the detector.

\begin{figure}[!t]
\centering
\includegraphics[width=1.5in]{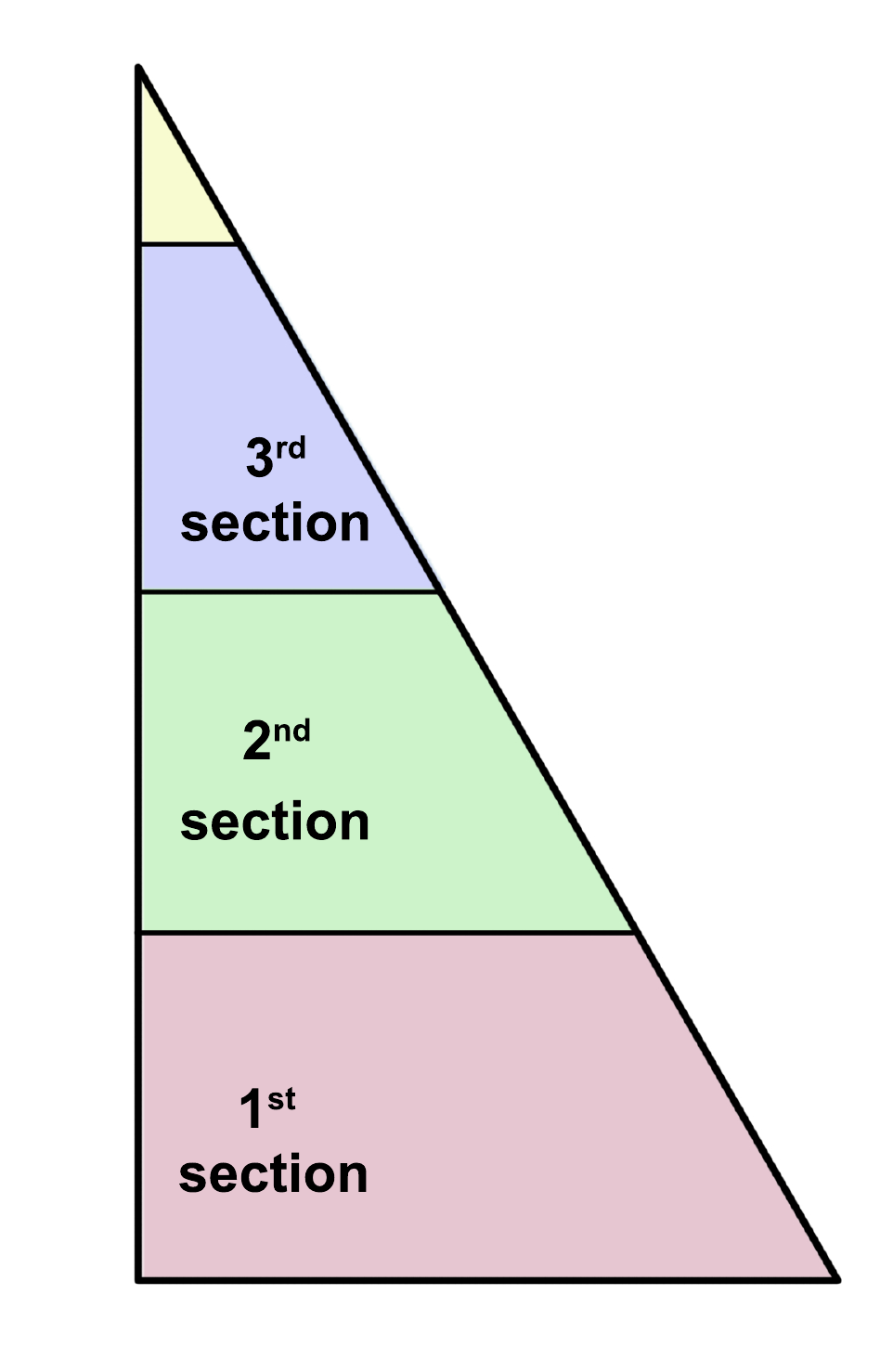}
\caption{An illustration of the side triangle of the sTGC prototype and its division into three trapezoidal strip groups.}
\label{fig:3_groups}
\end{figure}

\begin{figure}[!t]
\centering
\subfigure[]{
  \includegraphics[width=3.5in]{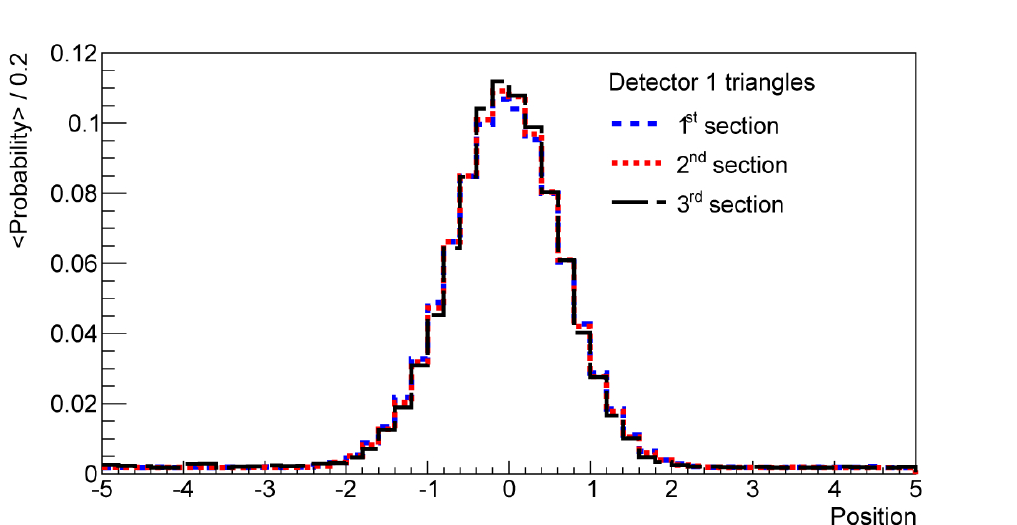}
}
\subfigure[]{
  \includegraphics[width=3.5in]{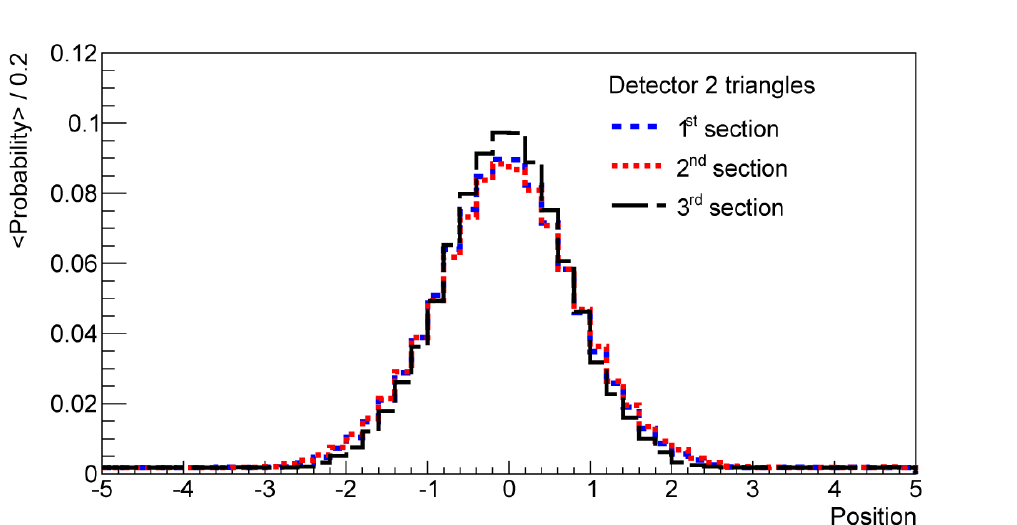}
}
\caption{The average area-normalized signal profile at the side triangles  in detector 1 (a) and detector 2 (b). The triangle was divided into three sections of 9-10 strips, such that section 1 has the largest area and section 3 has the smallest area.}
\label{fig:triangle_wide_dist}
\end{figure}

Hence, the signals near the detector edges are either narrower or partially lost.
In the latter case, the signal's shape would become asymmetric near the edge.
To examine this possibility, the triangles were divided into three sections of 9-10 strips each.
Each section forms a small trapezoid with different bases lengths, such that more wires are contained within the edge at sections with narrower bases, as illustrated in Figure~\ref{fig:3_groups}.
The average area-normalized profile of the signal in each section is shown in Figure~\ref{fig:triangle_wide_dist}.
As one can see, the signals are indeed becoming narrower as they approach the edge.
To check if edge-signals become asymmetrical or not we define a figure of merit that reflects such an asymmetric shape by the difference between the integral of the negative side of a signal profile of section $i$ in one of the triangles $f^i (x)$ and its positive side:

\begin{equation*}
A^{i}_{\Delta} = \sum_{x < 0} f^i (x) - \sum_{x > 0} f^i (x),
\end{equation*}
where x is the bin position.
The asymmetry of the central rectangle, $A_{\Box}$, is similarly defined.

The relative deviation of the asymmetry in each triangle section from the central rectangle, $A^{i}_{rel}$, in units of standard deviations is defined as:

\begin{equation*}
A^{i}_{rel} = \frac{A^{i}_{\Delta} - A_{\Box}}{\sigma_{\Box}},
\end{equation*}
where $\sigma_{\Box}$ is the standard deviation from the symmetric signal profile in the rectangle. 

The relative deviations of the asymmetry are shown in Table~\ref{tab2}.
In each case the difference is less than one standard deviation.
This implies that the signals remain symmetric even as the point of impingement approaches the edge, which means they are not cut (or partially lost).

\begin{table}[!t]
\begin{center}
\begin{tabular}{c|cccc}\hline\hline
\rule{0pt}{2ex} Section   &     $A_{rel}$ -- Detector 1      & $A_{rel}$ -- Detector 2\\\hline
\rule{0pt}{2ex} $1^{st}$    &    0.60                            &    0.42\\[1ex]
\rule{0pt}{2ex} $2^{nd}$  &     0.12                          &    -0.33\\[1ex]
\rule{0pt}{2ex} $3^{rd}$   &    0.85                          &    0.18\\
\hline\hline
\end{tabular}
\end{center}
\caption{The relative deviation of the asymmetry in each triangle section from the central rectangle. In each case the difference is less than $1\sigma$.}  
\label{tab2}
\end{table}

\section{Conclusions}\label{sec:TGC_concl}
One of the differences between the newly designed variant of the TGC chambers (the sTGCs) and the current ones is that the wires in the sTGC are approximately aligned along the azimuthal direction, perpendicular to the trapezoid bases.
As a result, the outermost wires approach the edge of the detector with a small angle.
Such a configuration is new and was suspected to give rise to unwanted effects.
In order to study the nature of such effects, a small prototype sTGC doublet was constructed and studied.

Despite the proximity of the wires to the edge, no sparks were observed during the tests, and the detectors work properly even when the operating voltage exceeds by 20\% its designed value.

However, signals near the edge of the detector appear to have lower amplitude and smaller charge. 
This effect was more significant in one detector than in the other. 
However, the difference between the two chambers of the doublet might be explained by differences in their construction.
Despite the differences between the detectors, the signals near the edges are consistently narrower by $\sim ~15\%$ than in the central parts of the chamber.
Yet, the signals remain symmetric even as they approach the edge.

While this study shows that the new sTGC chambers are robust and are not likely to spark, it reveals the existence of an edge effect.
the impact of this effect on the spatial resolution and efficiency requires a further study.
The fact that both chambers consistently show symmetrical signals, even as the point of impingement approaches the edges, might imply that the spatial resolution will not suffer from a systematic bias.
Full size prototypes with small angle wire edges were recently built at the Weizmann Institute, and will be used to further study the spatial resolution and other effects near the edge.

\bibliographystyle{IEEEtran}
\bibliography{IEEEabrv,ANIMMA}
%
%
%
%

\end{document}